\documentclass[11pt]{article}
\usepackage{hyperref}
\usepackage{amsmath, amsfonts, amssymb, mathrsfs}
\usepackage{dcolumn}
\usepackage{caption}
\usepackage{subcaption}
\usepackage{filemod}
\usepackage{natbib}
\usepackage[ruled, vlined]{algorithm2e}
\usepackage{floatrow}
\usepackage{setspace}
\usepackage{verbatim}
\usepackage{graphicx}

\usepackage{changes} 

\makeatletter
\@namedef{Changes@AuthorColor}{black}
\colorlet{Changes@Color}{black}
\makeatother

\hypersetup{
    colorlinks=true,
    linkcolor=blue, 
    urlcolor=black,
    citecolor=blue, 
    }

\title{\vspace{-0.3cm} Vecchia-approximated Deep Gaussian Processes\\for Computer Experiments}
\author{Annie Sauer\thanks{Corresponding author: Department of Statistics, 
	Virginia Tech, {\tt anniees@vt.edu}} 
	\and Andrew Cooper\thanks{Department of Statistics, Virginia Tech} 
	\and Robert B. Gramacy\footnotemark[2]}
\date{\today}

\begin{document}

\maketitle


\vspace{-0.3cm}
\begin{abstract} 
Deep Gaussian processes (DGPs) upgrade ordinary GPs through functional
composition, in which intermediate GP layers warp the original inputs,
providing flexibility to model non-stationary dynamics. Two DGP regimes have
emerged in recent literature.  A ``big data'' regime, prevalent in machine
learning, favors approximate, optimization-based inference for fast,
high-fidelity prediction. A ``small data'' regime, preferred for computer
surrogate modeling, deploys posterior integration for enhanced
uncertainty quantification (UQ). We aim to bridge this gap by expanding the
capabilities of Bayesian DGP posterior inference through the
incorporation of the Vecchia approximation, allowing linear computational
scaling without compromising accuracy or UQ. We are motivated by surrogate
modeling of simulation campaigns with upwards of 100,000 runs --
a size too large for previous fully-Bayesian implementations -- and
demonstrate prediction and UQ superior to that of ``big data'' competitors.
All methods are implemented in the {\tt deepgp} package on CRAN.
\end{abstract}



\section{Introduction}\label{sec:1}

Virtualization and simulation increasingly play a fundamental role in the
design and study of complex systems that are either impossible or infeasible 
to experiment with directly. Examples abound
in engineering \cite[e.g.,][]{zhang2015microstructure},
aeronautics \citep[e.g.,][]{mehta2014modeling}, economics
\citep[e.g.,][]{kita2016realistic}, and ecology
\citep[e.g.,][]{johnson2008}, to name a few.  Simulation is ``cheaper'' than
physical experimentation, but not ``free''. Associated computational costs
often necessitate statistical surrogates, meta-models that furnish accurate 
predictions and appropriate uncertainty quantification (UQ) from a limited 
simulation campaign. Surrogates
may thus stand-in for novel simulation to support downstream tasks such as
calibration \citep{kennedy2001bayesian}, optimization
\citep{jones1998efficient}, and sensitivity analysis
\citep{marrel2009calculations}.  Surrogate accuracy, combined with effective 
UQ, is key to the success of such enterprises.

Gaussian processes (GPs) are a common surrogate modeling choice
\citep{rasmussen2005gaussian,gramacy2020surrogates} because they offer both
accuracy and UQ in a semi-analytic nonparametric framework.  However,
inference for GPs requires the evaluation of multivariate normal (MVN)
likelihoods, which involves dense matrix decompositions that scale cubically
with the size of the training data. Computer simulation campaigns used to be
small, but recent advances in hardware, numerical libraries, and STEM training
have democratized simulation and led to massively larger campaigns
\citep[e.g.,][]{marmin2022deep,kaufman2011efficient,lin2021uncertainty,
sun2019emulating,liu2017dimension}. The literature has since adapted to this
bottleneck by borrowing ideas from machine learning and geo-spatial GP
approximation. Examples include sparse kernels \citep{melkumyan2009sparse},
local approximations
\citep{emery2009kriging,gramacy2015local,cole2021locally}, inducing points
\citep{quinonero2005unifying,finley2009improving}, and random feature
expansions \citep{lazaro2010sparse}. See \cite{heaton2019case} and
\cite{liu2020gaussian} for thorough reviews. Here we are drawn to a family of
methods that leverage ``Vecchia'' approximation \citep{vecchia1988estimation},
which imposes a structure that generates a sparse Cholesky factorization of
the precision matrix \citep{katzfuss2020vecchia,datta2021sparse}. When
appropriately scaled or generalized
\citep{stein2004approximating,stroud2017bayesian,datta2016hierarchical,katzfuss2021general},
and matched with sparse-matrix and multi-core computing facilities,
Vecchia-GPs dominate competitors \citep{katzfuss2020scaled} on
the frontier of accuracy, UQ, and speed.
  
Despite their prowess in many settings, GPs (and many of their approximations)
are limited by the assumption of stationarity; they are not able to identify
spatial changes in input--output regimes or abrupt shifts in dynamics.  Here
too, myriad remedies have been suggested across disparate literatures.  Common
themes include partitioning
\citep[e.g.,][]{kim2005analyzing,gramacy2008bayesian,rushdi2017vps,park2018patchwork},
evolving kernels
\citep{paciorek2003nonstationary,picheny2013nonstationary},
\added{and injective warpings \citep{zammit2021deep}}. Deep Gaussian
processes (DGPs) -- originating in geo-spatial communities
\citep{sampson1992nonparametric,schmidt2003bayesian} but recently popularized
by \cite{damianou2013deep} with analogy to deep neural networks -- address
this stationary limitation by layering GPs as functional compositions. Inputs
are fed through intermediate Gaussian layers, \added{which act as 
``warped'' versions of the original inputs}, before reaching the response.
The structure of a DGP is equivalent to that of a ``linked GP''
\citep{ming2021linked}, except that middle layers remain unobserved.  When the
data generating mechanism is non-stationary, DGPs offer great promise. 
When regimes change abruptly, DGPs mimic partition schemes.
When dynamics change more smoothly, they mimic kernel evolution.  When
stationary, they gracefully revert to ordinary GP dynamics via identity
warpings.

Or at least that's the sales pitch. In practice things are more murky because
non-stationary flexibility is intimately twinned with training data size and
structure.  Unless the experiment can be designed to squarely target locations
of regime change \citep{sauer2021active}, a large training data set is needed
before these dynamics can be identified. Cubic scaling in flops is present in
multitude, with large dense matrices at each latent DGP layer.  Moreover, the
unobserved/latent layers pose a challenge for statistical inference as they
may not be analytically marginialized {\em a posteriori}.  Markov chain Monte
Carlo (MCMC) sampling \citep{sauer2021active,ming2021deep} exacerbates cubic
bottlenecks and limits training data sizes to the hundreds. Approximate
variational inference (VI) offers a thriftier alternative to full posterior
integration \citep{damianou2013deep,salimbeni2017doubly}, but at the expense
of UQ.  VI for DGPs does not in-and-of-itself circumvent cubic expense, but,
when coupled with inducing points and mini-batching \citep{ding2021sparse}, 
can be scaled up; allow
us to defer a thorough review to Section \ref{sec:4}. We find these approaches
to be ill-suited to our computer surrogate modeling setting, in which
signal-to-noise ratios are high and UQ is a must. We speculate this is because
those libraries target benchmark machine learning examples for classification,
or for regression settings with high noise.

In this work we expand the utility and reach of fully Bayesian posterior
inference for DGP surrogate models through Vecchia approximation at each
Gaussian layer.  We explore strategic updating of the so-called ``neighborhood
sets'' involved, based on the warpings of intermediate layers. With careful
consideration of the myriad specifications in this hybrid context, we
demonstrate that a Vecchia-DGP need not sacrifice predictive power nor
UQ compared to a full (un-approximated) analog. We also show that it
outperforms approximate DGP competitors in both accuracy and UQ on data sizes
up to 100,000. An open-source implementation is provided in the {\tt deepgp}
package on CRAN \citep{deepgp}.

The remainder of the paper is organized as follows. Section \ref{sec:2}
establishes notation while reviewing GPs, DGPs, and Vecchia approximation.
Section \ref{sec:3} details our Vecchia-DGP framework.  Section
\ref{sec:4} discusses implementation and reviews related work on/software for
large scale DGP inference with an eye toward contrast and to set up benchmark
exercises presented in Section \ref{sec:5}.  That section concludes with a
study of a real-world satellite drag computer experiment.  A brief discussion
is offered in Section \ref{sec:6}.

\section{Review of major themes}\label{sec:2}

\subsection{Gaussian processes: shallow and deep}\label{sec:2gp}

Let $f: \mathbb{R}^d \rightarrow \mathbb{R}$ represent a (possibly noisy)
black-box function, say to abstract a computer model simulation.  Consider
inputs $X$ of size $n\times d$ and corresponding outputs/observations $Y =
f(X)$ of size $n\times 1$.  Throughout we use lowercase $x_i$ to refer to the
$i^\mathrm{th}$ (transposed) row of $X$, and likewise for $Y$.  Generic GP 
regression assumes a MVN over the response, $Y\sim
\mathcal{N}_n\left(\mu(X), \Sigma(X)\right)$. We specify $\mu(X) = 0$, as is
common in surrogate modeling \citep{gramacy2020surrogates}, yielding the 
likelihood
\begin{equation}\label{eq:likelihood}
\mathcal{L}(Y\mid X) \propto |\Sigma(X)|^{-1/2} 
	\exp \left(-\frac{1}{2}Y^\top \Sigma(X)^{-1} Y\right).
\end{equation}
Above ``$\propto$'' indicates a dropped multiplicative constant. 
All of the action is in $\Sigma(X)$, which is usually specified 
pairwise as a function of Euclidean distance between rows of $X$:
\begin{equation}\label{eq:sigma}
\Sigma(X)^{ij} = \Sigma(x_i, x_j) = 
\tau^2\left(k\left(\frac{||x_i - x_j||^2}{\theta}\right) 
+ g\mathbb{I}_{i = j}\right).
\end{equation}
Any choice of {\em kernel} $k(\cdot)$ leading to positive definite $\Sigma(X)$
is valid.  Most often these are exponentially decreasing in their argument.
Popular choices include the squared exponential and Mat\`ern
\citep{stein1999interpolation}, but our work here is not kernel specific.
Hyperparameters $\tau^2$, $\theta$, and $g$ govern the scale, lengthscale, and
noise respectively, working together to describe signal-to-noise
relationships.

There are many variations on this theme.  For example, vectorized $\theta$
allow the rate of decay of correlation to vary with input direction.
Such embellishments implement a form of {\em anisotropy}, a term preferred in
geo-spatial contexts, or {\em automatic relevance determination} in machine
learning \citep{liu2020}. We do not need such modifications in our DGP setup;
once latent layers are involved, such flexibility manifests more parsimoniously 
via those values.  Several of our competitors [Section \ref{sec:5}] do utilize 
additional hyperparameters and/or equivalent affine pre-scaling 
\citep{wycoff2021sensitivity}.

Settings of $(\tau^2, \theta, g)$ may be fitted through the likelihood
(\ref{eq:likelihood}), now viewing $(X,Y)$ as training data.  While derivative-based
maximization is possible with numerical solvers, we prefer full posterior
inference. We usually fix $g$ at a small constant, e.g., $g=10^{-6}$, as
is appropriate for deterministic (or very low noise) computer model
simulations.  However, we don't see this as a limitation of our contribution,
and our software allows for this parameter to be inferred if needed. Further
discussion is reserved for Section \ref{sec:6}. Regardless of these
choices, evaluation of the likelihood (\ref{eq:likelihood}) relies on both the
inverse and determinant of $\Sigma(X)$.  For a dense $n\times n$ matrix, 
this is an $\mathcal{O}(n^3)$ operation with conventional libraries.

Conditioned on training data and covariance formulation $\Sigma(\cdot)$, 
predictions at testing locations $\mathcal{X}$ (of size $n_p\times d$) 
follow Eq.~(\ref{eq:gppred}) after extending $\Sigma(\mathcal{X}, X)$ to 
cover rows of $\mathcal{X}$ paired with $X$ following Eq.~(\ref{eq:sigma}): 
\begin{equation}\label{eq:gppred}
\mathcal{Y} \mid Y, X \sim \mathcal{N}_{n_p}\left(\mu', \Sigma'\right) 
\;\;\textrm{where}\;\; 
\mu' = \Sigma(\mathcal{X}, X) \Sigma(X)^{-1} Y
\;\;\textrm{and}\;\;
\Sigma' = \Sigma(\mathcal{X}) - \Sigma(\mathcal{X}, X) 
	\Sigma{}(X)^{-1} \Sigma(X, \mathcal{X}).
\end{equation}
This closed-form is convenient but requires the inverse of $\Sigma(X)$, or at
least a clever linear solve.

The above GP specification is {\em stationary} because only relative positions of
training and testing inputs are involved (\ref{eq:sigma}).  Consequently,
identical input--output dynamics apply everywhere in the input space, which
can be limiting for some computer simulations.  A great example comes
from aeronautics/fluid dynamics.  Lift forces on aircraft are
fundamentally different at high speed versus low, and the transition at
the sound barrier is abrupt \citep{pamadi2004aerodynamic}.
Several strategies to relax stationarity were introduced in Section
\ref{sec:1}. Here we consider deep Gaussian processes
\citep[DGP;][]{damianou2013deep} which non-linearly warp inputs into a
plausibly stationary regime.  DGPs are functional compositions of GPs -- the
outputs of one GP feed as inputs to another. These intermediate Gaussian
layers combine to bring some input locations closer together while spacing
others further apart; \added{a visual will be provided momentarily}. While DGPs 
may extend several layers deep, we restrict
our discussion here to two layers for simplicity.  No part of our contribution
(nor software) is strictly limited to this choice, although there
is empirical evidence of diminishing returns for deeper DGP surrogates
\citep{rajaram2021empirical,sauer2021active}.

A two-layer DGP with latent layer $W$ is modeled as
\begin{equation}
\begin{aligned}
Y \mid W &\sim \mathcal{N}_n \left(0, \Sigma(W)\right), & \quad\quad
W_k &\stackrel{\mathrm{ind}}{\sim} \mathcal{N}_n \left(0, \Sigma_k(X) \right), 
	\quad k = 1,\dots,p.  \label{eq:dgp}
\end{aligned}
\end{equation}
$W = [W_1, \dots, W_p]$ is an $n\times p$ matrix with each row representing
one observation and each column representing a dimension of the latent space.
We adopt the deep learning convention of referring to the component dimensions
of $W$ as ``nodes''.  Each node has its own $\Sigma_k(X)$ which may or may
not share features, like kernels and hyperparameters, with others. In our
setting we specify unit scale and zero noise on latent layers (i.e., $\tau^2 =
1$ and $g = 0$ within $\Sigma_k(X)$) in order to preserve parsimony and
identifiability. It is common to fix $p = d$, but autoencoding setups which
``squeeze through'' lower-dimensional latent layers ($p \ll d$) are not
uncommon in high-dimensional settings \citep{domingues2018deep}. Again, our
work and implementation are not limited to this choice.

Inference for DGPs, including for hyperparameters buried in the $\Sigma$s, 
requires integration over $W$:
\begin{equation}\label{eq:marg}
\mathcal{L}\left(Y\mid X\right) = \int \mathcal{L}\left(Y\mid W\right)
\prod_{k=1}^p 
\mathcal{L}\left(W_k\mid X\right)\; dW,
\end{equation}
with $\mathcal{L}(Y\mid W)$ and $\mathcal{L}(W_k\mid X)$ following
Eq.~(\ref{eq:likelihood}), or slight variations thereupon.  However, this integral
is not analytically tractable. The prevailing inferential tools thus 
rely either on approximate variational methods
\citep{damianou2013deep,salimbeni2017doubly} or sampling
\citep{dunlop2018deep,havasi2018inference,ming2021deep}.

Our preferred inferential scheme, detailed in \cite{sauer2021active},
prioritizes UQ through a fully-Bayesian MCMC algorithm, hinging on elliptical
slice sampling \citep[ESS;][]{murray2010elliptical} of latent layers. ESS is
specifically designed for sampling variables with MVN priors, is notably free 
of tuning parameters, and works well in DGP settings 
\citep[also see][]{ming2021deep}.  We further embrace a Bayesian 
treatment of kernel hyperparameters: marginializing
$\tau^2$ from the posterior under a reference prior \citep[][Chapter
5]{gramacy2020surrogates} and adopting Metropolis-Hastings
\citep[MH;][]{hastings1970monte} sampling of $\theta$'s (each $\Sigma(W)$ and
$\Sigma_k(X)$ with unique lengthscale),
all wrapped in a Gibbs scheme.  This approach is thorough but computationally
demanding.  Each Gibbs iteration requires many evaluations of Gaussian
likelihoods (\ref{eq:likelihood}), and several thousands of iterations are
needed to burn-in and then explore the posterior.

\begin{figure}[ht!]
\centering
\includegraphics[width=17cm,trim=10 10 0 20]{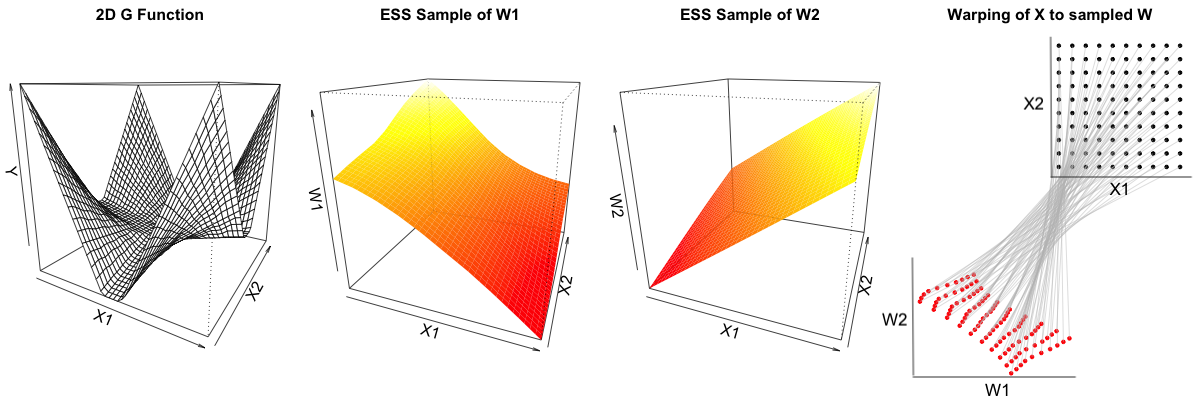}
\caption{(Left) Visual of the 2d G-function. (Middle Left/Middle Right)
Posterior sample of $W$ for a  two-layer DGP fit to the data of the left
panel, comprised of two ``nodes'' (one in each panel). \added{(Right) The resulting
``warping'' of this sample as evenly gridded $X$ are mapped to $W$.}}
\label{fig:g}
\end{figure}

As an example of DGP warping, consider the two-dimensional ``G-function'',
featured later in Section \ref{sec:5sims}.  A visual is provided in the left
panel of Figure \ref{fig:g}.  This function is characterized by inclines and
abrupt shifts. A stationary ``shallow'' GP is unable to distinguish between
the steep sloping regions in the corners and the valley regions that form a
cross shape in the center. Figure \ref{fig:g} also shows a posterior sample
(i.e., a burned-in ESS draw) of the hidden layer $W$ plotted \added{two ways.
The middle panels show each of two ``nodes'' as a function of $X$.  Each node
is a conditionally independent stationary GP.  These nodes depart from the
identity mapping (i.e. $W_1 = X_1$ and $W_2 = X_2$) by stretching the inputs
along the diagonal. The right panel shows the combined ``warping'' effect that
both nodes have on the evenly gridded $X$. Since only pair-wise distances feed
into the final layer (through $\Sigma(W)$), the warping effect is invariant to
rotations and translations.  ESS samples of $W$ that ``bounce''  between
mirror images of themselves are taken as an indication that the chain is mixing
well.  In Section \ref{sec:5sims} we show that the warping(s) in the figure
accommodate the G-function's non-stationary dynamics better than many other
methods.}

\subsection{Vecchia approximation}\label{sec:2vec}

The Vecchia approximation \citep{vecchia1988estimation} is motivated by
computational bottlenecks in GP regression, which are compounded in a 
DGP setting. The underlying idea is basic: 
any joint distribution can be 
factored into a product of conditionals $p(y) = p(y_1) p(y_2 \mid y_1) 
p(y_3 \mid y_2, y_1) \cdots p(y_n \mid y_{n-1}, \dots, y_1)$.  This is true 
up to any re-indexing of the $y_i$'s.  In particular, and to establish
some notation for later, any joint likelihood (\ref{eq:likelihood}) may be
factored into a product of univariate likelihoods
\begin{equation}\label{eq:approx}
\mathcal{L}\left(Y\right) = 
	\prod_{i=1}^n \mathcal{L}\left(y_i \mid Y_{c(i)}\right)
\end{equation}
where $c(1) = \emptyset$ and $c(i) = \{1, 2, \dots, i - 1\}$ for $i = 2,
\dots, n$.  The Vecchia approximation instead takes a subset, $c(i) \subset
\{1, 2, \dots, i - 1\}$, of size $|c(i)| = \min\left(m, i - 1\right)$.  When
$m < n$, the strict equality of Eq.~(\ref{eq:approx}) is technically an
approximation, yet we will use equality notation throughout when speaking of
the general case with unspecified $m$.  This approximation is
indexing-dependent for fixed $m < n$, but hold that thought for a moment. 
Crucially Eq.~(\ref{eq:approx}), in the context of Eqs.~(\ref{eq:likelihood})
and (\ref{eq:gppred}), induces a sparse precision matrix: $Q(X) =
\Sigma(X)^{-1}$.  The $(i, j)^\mathrm{th}$ element of $Q(X)$ is 0 if and only
if $y_i$ and $y_j$ are conditionally independent (i.e. $i \notin c(j)$ and $j
\notin c(i)$). The Cholesky decomposition of the precision matrix, $U_x$ for
$Q(X) = U_xU_x^\top$, is even sparser with fewer than $m$ off-diagonal
non-zero entries in each row.  We follow \cite{katzfuss2020vecchia} in
working with the upper trianglular $U_x$, referred to as the ``upper-lower'' Cholesky
decomposition.

\begin{figure}[ht!]
\centering
\includegraphics[width=18cm,trim=10 15 0 10]{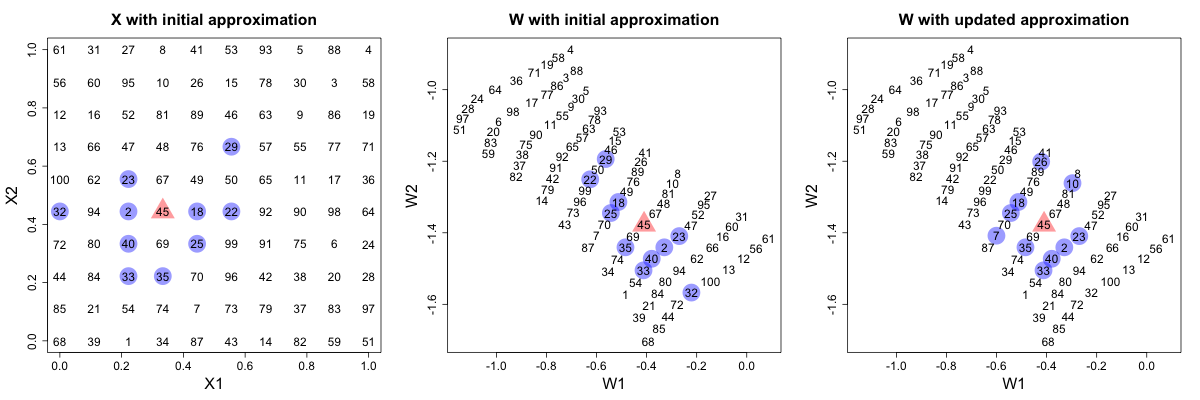}
\caption{(Left) A uniformly spaced grid of inputs, randomly ordered.
NN conditioning set for $x_{45}$ (triangle) is highlighted by
circles.  \added{(Middle) Same ordering/NN sets instead plotted as the ``warped'' $W$
from Figure \ref{fig:g}.  (Right) ``Warped'' $W$ with NN 
conditioning adjusted accordingly.}}
\label{fig:g_order}
\end{figure}

A GP-Vecchia approximation requires two choices: an ordering of the data 
and selection of conditioning sets $c(i)$.  There are many orderings that work
well \citep{stein2004approximating,guinness2018permutation,katzfuss2021general},
but a simple random ordering is common
\citep{stroud2017bayesian,datta2016hierarchical,wu2022variational}.  The
prevailing choice for conditioning sets is ``nearest-neighbors'' (NN) in which
$c(i)$ comprises of integers indexing the closest observations to $x_i$
which appear earlier in the ordering.  Approximations based on NN are
sometimes referred to as NNGPs \citep{datta2016hierarchical}. To demonstrate
an ordering and NN conditioning set, the left panel of Figure
\ref{fig:g_order} shows a grid of inputs with random ordering (the numbers
plotted).  [The other panels will be discussed in Section \ref{sec:3}.] For
point $i = 45$ (triangle), the NN conditioning set of size $m
= 10$ is highlighted by circles.  These are the points closest to
$x_{45}$ in Euclidean distance, with indices $j < i$ in the ordering. Sets
$c(i)$ for all $i = 1, \dots, 100$ are chosen similarly.

Under a GP, components of Eq.~(\ref{eq:approx}) are univariate Gaussian,
$\mathcal{L}(y_i \mid Y_{c(i)}) \sim 
	\mathcal{N}_1(\mu_i(X), \sigma_i^2(X))$, where
\begin{equation}\label{eq:vecdirect}
B_i(X) = \Sigma(x_i, X_{c(i)})\Sigma(X_{c(i)})^{-1}, \quad
\mu_i(X) = B_i(X) Y_{c(i)}, \quad
\sigma_i^2(X) = \Sigma(x_i) - B_i(X)\Sigma(X_{c(i)}, x_i)
\end{equation}
and $X_{c(i)}$ is the row-combined matrix of $X$'s rows corresponding to
indices $c(i)$.  Foreshadowing a DGP application in Section \ref{sec:3}, one
may define $B_i(W)$, $\mu_i(W)$ and $\sigma_i^2(W)$ identically but with $w/W$
in place of $x/X$.  With this representation, we convert a large
$n\times n$ matrix inversion ($\mathcal{O}(n^3)$) into $n$-many $m\times m$
matrix inversions ($\mathcal{O}(nm^3)$), a significant improvement if $m \ll
n$.

The details of Vecchia-GPs, including numerous options for orderings,
conditioning sets, hyperparameterizations, and computational considerations,
are spread across multiple works \citep[e.g.,][]{katzfuss2020vecchia,
katzfuss2021general,guinness2018permutation,
guinness2021gaussian,datta2016hierarchical,datta2021sparse,finley2019efficient}.
These specifications, along with software implementations
\citep[e.g.][]{GPvecchia,GpGp,spNNGP}, can be rather complex.  We do not need
such hefty machinery in our DGP setup for computer surrogate models.  Much of
our research effort involved sifting through this literature to determine what
is essential and which variations work best for DGP surrogates.  As one
example, latent layers $W$ and deterministic $Y=f(\cdot)$ utilize noise free
modeling (small/zero nugget $g$), which affords several simplifications.  In
particular, we do not follow \cite{datta2016hierarchical} and
\cite{katzfuss2021general} in distinguishing between noisy observations and
latent ``true'' variables. This greatly streamlines development [Section
\ref{sec:3}] and reduces computational demands.

\section{Vecchia-approximated deep Gaussian processes}\label{sec:3}

Here we detail our Vecchia-DGP model, posterior integration, and other implementation
considerations. Our focus remains on two-layer models for simplicity; extension to
deeper DGPs comes down to iteration.  We begin by fixing ordering and
conditioning; ideas to tailor these to the DGP context follow later in Section
\ref{sec:3ordering}. Our software implementation [Section
\ref{sec:4software}], supports a wider array of options than those enumerated
here, including deeper (Vecchia) DGPs and estimating nuggets for smoothing
noisy data [Section \ref{sec:6}].

\subsection{Inferential building blocks}\label{sec:3blocks}

We impose the Vecchia approximation at each layer of the DGP.  Leveraging
sparsity of the upper-lower Cholesky decomposition of the precision matrices,
a two-layer Vecchia-DGP model may be represented as
\begin{equation}\label{eq:dgpvec}
\begin{aligned}
Y \mid W &\sim \mathcal{N}_n\left(0, (U_wU_w^\top)^{-1} \right)  
& \quad\quad
W_k &\stackrel{\mathrm{ind}}{\sim} \mathcal{N}_n\left(0,
	\left((U_x^{(k)})(U_x^{(k)})^\top\right)^{-1}\right), \quad k = 1,\dots,p,
\end{aligned}
\end{equation}
where $\Sigma(W)^{-1} = U_wU_w^\top$ and $\Sigma_k(X)^{-1} =
(U_x^{(k)})(U_x^{(k)})^\top$. Each $W_k$, having its
own Gaussian prior, {\em also} has its own Vecchia decomposition. Often these
$U_x^{(k)}$ will share conditioning sets but have disparate
hyperparameterization, e.g., unique lengthscales $\theta_{x}^{(k)}$.  
When $m= n$, this formulation is equivalent to Eq.~(\ref{eq:dgp}). When 
$m < n$, $U_w$ and $U_x^{(k)}$ have induced sparsity.

We aim to conduct full posterior integration for this model, extending
Eq.~(\ref{eq:marg}) to include integration over hyperparameters, but, as in Section
\ref{sec:2gp}, this is not analytically tractable.  Posterior sampling via
ESS and MH regarding $\mathcal{L}(Y\mid W)$ and $\mathcal{L}(W_k\mid X)$
requires three ingredients: (i) prior sampling, (ii) likelihood evaluation,
and (iii) prediction at unobserved inputs. These are detailed here, for the
model in Eq.~(\ref{eq:dgpvec}), in turn.  We shall focus on $\mathcal{L}(Y\mid
W)$, but the idea is immediately extendable to $\mathcal{L}(W_k\mid X)$. Both
are GPs, so the details only differ superficially in notation, and with
iteration over $k=1,\dots,p$. Ultimately, these ``building blocks'' tie
together to support posterior sampling, with details following in Section
\ref{sec:3inference}.

\paragraph{(i) Prior.} Direct sampling of 
$Y^\star \sim \mathcal{N}_n(0, (U_wU_w^\top)^{-1})$, often called
``conditional simulation'', involves individually drawing $y_i^\star \sim
\mathcal{N}_1(B_i(W)Y_{c(i)}^\star, \sigma_i^2(W))$ for $B_i(W)$
and $\sigma_i^2(W)$ defined with ana{}logy to Eq.~(\ref{eq:vecdirect}). An
important difference, however, is that the application here crucially relies
on {\it previously sampled} $Y_{c(i)}^\star$, meaning each $y_i^\star$ must be
sampled {\it sequentially}.  With an eye towards parallel implementation
[Section \ref{sec:4software}], we instead leverage the sparsity of the
upper-triangular $U_w$.  \citet[][Proposition 1]{katzfuss2021general} derived a
closed-form solution for populating $U_w$, with $(j, i)^\mathrm{th}$ entry
\begin{equation}\label{eq:u}
U_w^{ji} = \begin{cases}
\frac{1}{\sigma_i(W)} & i = j \\
-\frac{1}{\sigma_i(W)} B_i(W)[\textrm{index of } j \in c(i)] & j \in c(i)\\
0 & \textrm{otherwise}
\end{cases}
\end{equation}
for $B_i(W)$ and $\sigma_i(W)$ in Eq.~(\ref{eq:vecdirect}).  With 
$U_w$ in hand, sampling $Y^\star$ follows \citet[][App.~A]{gelman2013bayesian}:
\begin{equation}\label{eq:prior} 
Y^\star = (U_w^\top)^{-1}z  \quad\textrm{where}\quad 
z \sim \mathcal{N}_n(0, \mathbb{I}).
\end{equation} 
Strategically, we avoid matrix inversions by using a forward
solve of $U_w^\top Y^\star = z$.

\paragraph{(ii) Likelihood.} Evaluations of $\mathcal{L}(Y\mid W)$ 
could similarly be calculated as the product 
of univariate Gaussian densities, combining 
Eqs.~(\ref{eq:approx}--\ref{eq:vecdirect}) via
$\mathcal{L}(y_i \mid W) \sim \mathcal{N}_1(\mu_i(W), \sigma_i^2(W))$.
We instead choose to leverage the sparse $U_w$ formulation (\ref{eq:u}),
yielding the log likelihood 
\begin{equation}\label{eq:veclik}
\begin{aligned}
\log\mathcal{L}(Y \mid W) &\propto \log|(U_wU_w^\top)^{-1}|^{-1/2} 
	- \frac{1}{2}Y^\top U_wU_w^\top Y \\
	&\propto \sum_{i=1}^n \log(U_w^{ii}) - \frac{1}{2}Y^\top 
		U_wU_w^\top Y, \\
\end{aligned}
\end{equation}
in which the sparse structure of $U_w$ allows for thrifty matrix multiplications.

\paragraph{(iii) Prediction.} Given observed $Y \mid W \sim 
\mathcal{N}_n(0, (U_wU_w^\top)^{-1})$, i.e., training observations
$Y$ and a burned-in ESS sample of $W$, we wish to predict $\mathcal{Y}$ for an
$n_p\times d$ matrix of novel $\mathcal{W}$.  Note these novel $\mathcal{W}$
ultimately arise as samples following analogous application of the very same
procedure we are about to describe, except for $W_k$ as the ``response'' drawn
at novel testing sites $\mathcal{X}$ (more on this in Section \ref{sec:3inference}).
The simplest approach treats each row of
$\mathcal{W}$ independently.  Independent prediction is sufficient if only
point-wise means and variances are required, as is common in many downstream
surrogate modeling tasks. For each $i = 1, \dots, n_p$, we form $c(i)$ with
$m$ training locations from $W$ (details in Section \ref{sec:3ordering}).
This imposes conditional independence among $\mathcal{Y}_i$ (i.e., $\mathcal{Y}_i$
is not conditioned on $\mathcal{Y}_j$ for $i \neq j$).  The posterior
predictive distribution then follows $\mathcal{Y}_i \sim
\mathcal{N}_1(\mu_i(W), \sigma_i^2(W))$ with $\mu_i(W)$ and $\sigma_i^2(W)$
defined as in Eq.~(\ref{eq:vecdirect}).

This independent treatment is fast and easily parallelized over
index $i = 1, \dots, n_p$.  Consequently, it is the method we prefer for the
benchmarking exercises of Section \ref{sec:5}, involving a cumbersome
additional layer of Monte Carlo (MC) over training--testing partitions. Yet an
imposition of independence among $\mathcal{Y}$ can be limiting. In some cases,
joint prediction utilizing the full covariance structure $\mathcal{Y}
\mid Y, W \sim \mathcal{N}_{n_p}(\mu^\star, \Sigma^\star)$ is essential. 
We can accommodate such settings as follows.   First append $\mathcal{W}$
indices to the existing ordering of $W$, ensuring predictive locations
are ordered {\it after} training locations, forming the full ordering $i = 1,
\dots, n, n+1, \dots, n + n_p$.   Conditioning sets $c(i)$ for 
$i = n + 1, \dots, n + p$ index {\it any} observations from $W$ or
$\mathcal{W}$ with indices prior to $i$ in the combined ordering, thus
allowing predictive outputs to potentially condition on other predictive
outputs, in addition to nearby training data observations.

Next, ``stack'' training and testing responses in the usual way \citep[][Section
5.1.1]{gramacy2020surrogates}
\[
\begin{bmatrix} Y \\ \mathcal{Y} \end{bmatrix}
\sim \mathcal{N}_{n + n_p} \left(0, \;\; \Sigma_{\textrm{stack}}\right)
\quad\textrm{where}\quad 
\Sigma_\textrm{stack} = \Sigma\left(\begin{bmatrix} W \\ \mathcal{W} \end{bmatrix}\right) = 
\begin{bmatrix} \Sigma(W) & \Sigma(W, \mathcal{W}) \\ 
	\Sigma(\mathcal{W}, W) & \Sigma(\mathcal{W}) \end{bmatrix}.\
\]
Then leverage (\ref{eq:u}) to analogously populate a ``stacked'' 
upper-lower Cholesky decomposition,
\[
U_{\textrm{stack}} = \begin{bmatrix} U_{w} & U_{w,\mathcal{W}} \\
	0 & U_{\mathcal{W}} \end{bmatrix}
\quad\textrm{such that}\quad
\Sigma_{\textrm{stack}} = \left(U_{\textrm{stack}} U_{\textrm{stack}}^\top\right)^{-1}
= \left(\begin{bmatrix}
U_wU_w^\top + U_{w,\mathcal{W}}U_{w,\mathcal{W}}^\top & U_{w,\mathcal{W}}U_\mathcal{W}^\top \\
U_\mathcal{W}U_{w,\mathcal{W}}^\top & U_\mathcal{W}U_\mathcal{W}^\top \\
\end{bmatrix}\right)^{-1}.
\]
An application of the partition matrix inverse identities (details in 
App.~\ref{app:inv}--\ref{app:pred}) results in the following posterior 
predictive moments, after applying the usual MVN conditioning identities 
for $\mathcal{Y} \mid Y, W$ (\ref{eq:gppred}):
\begin{equation}\label{eq:vecpred}
\mathcal{Y} \mid Y, W \sim \mathcal{N}_{n_p}\left(\mu^\star, \Sigma^\star \right) 
\quad\textrm{for}\quad 
\mu^\star = -(U_\mathcal{W}^\top)^{-1} U_{w,\mathcal{W}}^\top Y
\quad\textrm{and}\quad
\Sigma^\star = \left(U_{\mathcal{W}} U_{\mathcal{W}}^\top\right)^{-1}.
\end{equation}
These are simplified versions of the moments provided by
\cite{katzfuss2020vecchia}, thanks to a streamlined latent structure and the
imposition that predictive locations must be ordered after training locations.
Naturally, if no predictive locations condition on others then
$U_\mathcal{W}$ and $\Sigma^\star$ will be diagonal, and we can return to the
simpler implementation of independent predictions. 

\citet{katzfuss2020vecchia} remark that conditioning on other predictive
locations, i.e., using joint $\mu^*$ and $\Sigma^\star$, is more accurate than
conditioning only on training data. 
Anecdotally, in our own empirical
work, we have found this difference to be inconsequential. Unless a joint
$\Sigma^\star$ is required, say for calibration \citep{kennedy2001bayesian},
we prefer the faster, parallelizable, independent approach. (Both are provided
by our software; more in Section \ref{sec:4software}.) In settings where it
might be desirable to reveal/leverage posterior predictive correlation,
but perhaps it is too computationally burdensome to work with 
$n_p^2$ pairs of testing sites simultaneously,  a hybrid or
batched scheme might be preferred.  

\subsection{Posterior inference}\label{sec:3inference}

Building blocks (i--iii) in hand, posterior sampling by MCMC may commence
following Algorithm 1 of \cite{sauer2021active}.  In other words, the
underlying inferential framework is unchanged modulo an efficient (Vecchia)
method for (i) prior sampling, (2) likelihood evaluation, and (3) prediction.
Rather than duplicate details here, allow us point out a few relevant
highlights. In model training, every evaluation of a Gaussian likelihood
utilizes Eq.~(\ref{eq:veclik}), whether for inner ($W_k$) or outer ($Y$),
matched with $U_x^{(k)}$ and $U_w$, respectively, and with appropriate
covariance hyperparameters, e.g., $\theta_x^{(k)}$, embedded into the
$B_i(\cdot)$ and $\sigma_i(\cdot)$ components of said $U$ matrices
(\ref{eq:u}). When employing ESS for $W^{(k)}$, say, random samples from the
prior follow Eq.~(\ref{eq:prior}). The MCMC scheme remains unchanged 
modulo Vecchia approximation everywhere under-the-hood.

To predict at unobserved inputs $\mathcal{X}$, i.e., Section 4.1 of
\cite{sauer2021active}, replace traditional GP prediction at each Gaussian
layer (\ref{eq:gppred}) with its Vecchia counterpart (\ref{eq:vecpred}).  For
each candidate (burned-in/thinned) MCMC iteration, predictive locations
$\mathcal{X}$ are mapped to ``warped'' locations $\mathcal{W}$, which are then
mapped to posterior moments for $\mathcal{Y}$, with each step following
Eq.~(\ref{eq:vecpred}). The resulting moments are post-processed, with ergodic
averages yielding the final posterior predictive moments.
\citet{sauer2021active} focused on small training and testing sets, so their
setup favored samples from a joint predictive distribution analogous to
Eq.~(\ref{eq:vecpred}). These may be replaced with independent point-wise, and
parallelized predictions as described in Section \ref{sec:3blocks} in the
presense of a large/dense testing set. We remind the reader that a DGP
predictive distribution is not strictly Gaussian, even though it arises as an
integral over Gaussians. However we find that ergodic
averages, represented abstractly here by empirical moments $\bar{\mu}$ and
$\bar{\Sigma} + \mathbb{C}\mathrm{ov}(\mu)$ through the law of total variance, are a
sufficient substitute for retaining thousands of high-dimensional MCMC draws
of $\mu^\star$ and $\Sigma^\star$, say, or their pointwise analogues.

It is important to briefly acknowledge the substantial computation inherent in
this inferential scheme.  The requisite MCMC requires thousands of iterations,
each of which necessitates multiple likelihood (\ref{eq:veclik}) evaluations.
Predictions require averaging across these draws, although thinning can reduce
this effort.  Despite these hefty computational demands, efficient
parallelization (described in more detail momentarily), strategic
initializations of latent layers, and other sensible pre-processing yield
feasible compute times even with large data sizes.  For example, the
Vecchia-DGPs of Section \ref{sec:5satellite} with $n = 100,000$ were fit
in 26 hours on a 16-core machine. 
We aim to show that this
investment pays dividends compared to faster GP and DGP alternatives in terms
of prediction accuracy and UQ [Section \ref{sec:5}].

\subsection{Ordering and conditioning}\label{sec:3ordering}

Each Gaussian component of the DGP could have its own ordering and
conditioning set, $c_x^{(k)}(i)$ for $U_x^{(k)}$ and $c_w(i)$ for $U_w$ in the
two-layer model, in which orderings denoted by $i$ need not be the same. Since
each $c_x^{(k)}(i)$ acts on the same input space, we simplify the
approximation by sharing ordering and conditioning sets across $k = 1, \dots,
p$, resulting in only two orderings and two conditioning sets, $c_x(i)$ and
$c_w(j)$.  

These choices are not part of the stochastic process describing the data
generating mechanism, although that is an interesting possibility we discuss
in Section \ref{sec:6}.  A consequence of this is that once a chain has been
initialized under a particular ordering, yielding $U_x$ or $U_w$ up to
hyperparameters like $\theta$ which {\em are} included in the hierarchy
describing the stochastic process, $c_x(i)$ and $c_w(j)$  must remain fixed
throughout the MCMC in order to maintain detailed balance. It occurred to us
to try randomizing over orderings from one MCMC iteration to the next, but the
chain does not burn in/achieve stationarity. Each change to one of $c_x(i)$ or
$c_w(j)$ causes the chain to ``jump'' somewhere else.  Nevertheless, it could
be advantageous to customize aspects of a Vecchia ordering and conditioning
dynamically, say based on a DGP fit or other analysis, or to hedge by
averaging results from multiple orderings. This is fine with independent
chains.

With this in mind, we adopt the following setup.  Begin with a random ordering
of indices for each DGP layer.  This follows the recommendation of
\cite{guinness2018permutation} and mirrors other recent work on NNGPs
\citep{wu2022variational,datta2016hierarchical,stroud2017bayesian}. Then
select conditioning sets based on NN, as eponymous in the NNGP/Vecchia
literature.  In $X$-space, calculating $c_x(i)$ as the $\min(m, i - 1)$ nearest
points (of lower index) is straightforward.  These locations are anchored in
place by the experimental design.  For latent Gaussian layer $W$, NN
conditioning sets can be more involved. Since $W$ is unobserved, we start with
no working knowledge of the nature of the warping. (Our prior is mean-zero
Gaussian under a distance-based covariance structure with unknown
hyperparameterization.)  A good automatic initialization for the MCMC is to
assume no warping (i.e. $W = X$). In that setting, NN for $W$ based on
relative Euclidean distance in $X$ space is sensible.

Such a conditioning set is even workable after considerable posterior
sampling, whereby $W$ may have diverged from the identity mapping with $X$. We
find that in practice each individual nodes' (i.e., $W_k$) contribution to the
overall multidimensional warping for $k=1,\dots,p$ is usually convex. As a
visual, consider again Figure \ref{fig:g_order}.  
\added{The right two panels show conditioning sets (both NN in a certain
sense) for $W$ arising as a function of $X$ corresponding to the 
mapping in Figure \ref{fig:g}.}
The locations of the observations (marked by
numbers) represent a warping of the original evenly-gridded inputs (left
panel).  The middle panel shows the conditioning set $c_w(45)$ that was
selected based on NN in $X$ space. Observe that these highlighted points are
equivalent to those of the left panel.

Selecting $c_w(j)$ based on $X$ is a good starting point, but we envision
scope to be more strategic. Given posterior information about $W$, one may
wish to update $c_w(j)$ in light of that warping.  For example, NN on $W$
could be calculated after burn-in and used as the basis of $U_w$ conditioning
for a re-started chain.  Such an operation could be viewed as a nonlinear
extension of sensitivity pre-warping \citet{wycoff2021sensitivity}, tailored
to the Vecchia approximation. The right panel of Figure \ref{fig:g_order}
shows \added{such a re-conditioning}. 
There is some precedence for evolving neighborhood sets in this
way from the ordinary Vecchia-GP literature.  For example,
\cite{katzfuss2020scaled} use estimated multiple-lengthscale parameters to
find NN based on re-scaled inputs $X/\sqrt{\theta}$, vectorizing over columns;
\cite{kang2021correlation} extend that to full kernel/correlation based
re-scaling. Both are situated in an optimization based inferential apparatus,
and the authors describe a careful ``epoch-oriented'' scheme to circumvent
convergence issues, analogous to maintaining detailed balance in MCMC. Our
particular re-burn-in instantiation of this idea, described above, represents
a natural extension: an affine warping of inputs for NN calculations is
upgraded to a nonlinear one via latent Gaussian layers.  Yet in our empirical
work exploring this idea [Section \ref{sec:5sims}], we disappointingly find
little additional value realized by the extra effort for DGPs.  We speculate
in Section \ref{sec:6} that this may be because we haven't yet encountered any
applications demanding highly non-convex $W$.

A final consideration involves extending orderings and neighborhood sets to
testing sites when sampling from the posterior predictive distribution. Here,
we again use NN sets to select $c_x(i)$ and $c_w(j)$ for predictive/testing
locations $\mathcal{X}$ (which are mapped to $\mathcal{W}$). In $X$ space, NN
sets are fixed once for each row of $\mathcal{X}$. In $W$ space, after MCMC
sampling (i.e., model ``training''), we leverage the learned warpings
($W^{(t)}$ for $t\in\mathcal{T}$)
and calculate NNs in that space to maximize the efficacy of the
approximation at each location: for each row of $\mathcal{W}$
we re-calculate ``warped'' NN sets for each sample $W^{(t)}$.  Resulting predictions are
combined with expectation taken over all $t\in\mathcal{T}$. This
re-calculation of $c_w(i)$ for each $t$ requires extra effort, but it is not
onerous and focuses computation where it is most needed. 

\section{Implementation and competition}\label{sec:4}

Here we detail our implementation and provide evidence of
substantial speedup compared to a full DGP utilizing the same underlying
method but without a Vecchia approximation \citep{sauer2021active}.  We see
this compartmentalization -- identical computation modulo sparsity of
inverse Cholesky factors -- as one of the great advantages of our approach.  To
explain and contrast, we then transition to a discussion of other DGP
variations where we find that engineering choices (for computational
efficiency) are far more strongly coupled to  modeling ones (for statistical
fidelity), which can adversely affect performance in surrogate modeling 
settings, i.e., high-signal/low noise prediction with appropriate UQ.

\subsection{Implementation}\label{sec:4software}

We provide an open-source implementation as an update to the {\tt deepgp}
package \citep{deepgp} for {\sf R} on CRAN.  Although we embrace a bare-bones
approach, our {\sf R/C++} implementations of the ``building blocks'' in
Section \ref{sec:3blocks} are heavily inspired by the more extensive {\tt
GPvecchia} \citep{GPvecchia} and {\tt GpGp} \citep{GpGp} packages.
Computational speed relies on strategic parallelization and careful
consideration of sparse matrices. For example, we utilize {\tt OpenMP} pragmas
to parallelize the calculation of each row of the sparse $U_w$ and $U_x^{(k)}$
(\ref{eq:u}).  We use {\tt RcppArmadillo} \citep{RcppArmadillo} in {\sc
C++} and {\tt Matrix} \citep{Matrix} in {\sf R} to handle sparse matrix
calculations, aspects of which are also parallelized (but usually to a lesser
degree) under-the-hood.

While our Vecchia-DGP implementation in {\tt deepgp} is distinct from its
full (un-approximated) counterpart, they share an interface for
ease of use.  A {\tt vecchia} indicator to the existing {\tt fit} functions
triggers approximate inference, \added{i.e.}
\begin{verbatim}
  R> fit <- fit_two_layer(x, y, vecchia = TRUE, m = 25, true_g = eps)
  R> fit <- predict(fit, x_pred, lite = TRUE, m = 25)
\end{verbatim}
The neighborhood size is specified 
as {\tt m} $= m = \min\{25,n-1\}$ by default, where choosing $n-1$ results in no
approximation, but still uses the Vecchia implementation, which is useful
for debugging and benchmarking.  \added{Notice that training and prediction 
accommodate distinct $m$.
We have found that $m=25$ (for both) works well for our exercises here 
[further investigation in Supp.~\ref{app:m}], but it may be 
advantageous to use larger $m$ for prediction or to scale $m$ with $n$.}
We additionally allow {\tt predict} calls to toggle between
independent ({\tt lite = TRUE}) or joint predictions ({\tt lite = FALSE}).

A two-phase MCMC option, updating orderings and NN conditioning sets based on
a burned-in warping [Section \ref{sec:3ordering}] is supported by
{\tt re\_approx = TRUE} to {\tt continue}, an S3 method automating
additional MCMC from the end of a previous chain.  The {\tt true\_g} argument
is optional.  If it is not provided, then a nugget is estimated along with
other hyperparameters.  In our experiments later we fix {\tt eps = 1e-8}
for all but the satellite drag example [Section \ref{sec:5satellite}] where
we follow others in using {\tt eps = 1e-4}.

\begin{figure}[ht!]
\centering
\includegraphics[width=18cm,trim=10 15 0 15]{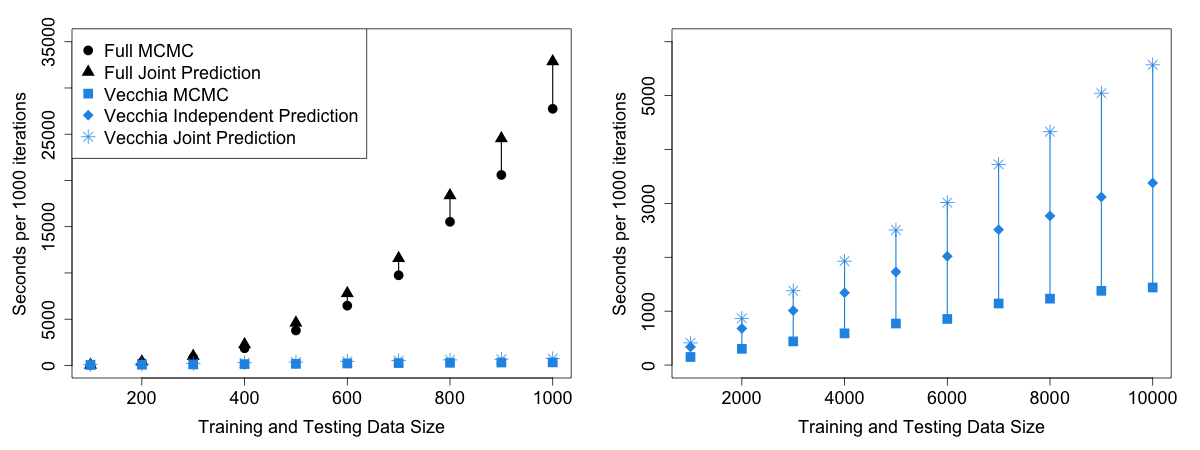}
\caption{\added{Computation time in seconds per 1000 MCMC iterations for a two-layer
DGP fit to the 2d Schaffer function.
Cumulative training and prediction times are connected by vertical lines.
Times are reported for a full un-approximated DGP (smaller 
data sizes) and a Vecchia-approximated DGP ($m = 25$).}}
\label{fig:time}
\end{figure}

To demonstrate computational improvements over the full un-approximated
implementation, the left panel of Figure \ref{fig:time} compares the
computation time of 1000 such MCMC iterations between the full implementation
of \cite{sauer2021active} and our proposed Vecchia-DGP.  A
16-core, hyperthreaded, Intel i9 CPU at 3.6GHz, was used to collect these
timings. The full, non-Vecchia implementation experiences cubic-in-$n$ costs
and is not generally feasible for sample sizes above several hundred.  The
Vecchia implementation scales linearly-in-$n$, allowing for much larger data
sizes.  To contrast approaches to prediction outlined in Section
\ref{sec:3blocks}, the right panel of Figure \ref{fig:time} shows computation
times for Vecchia-DGPs on larger training data sizes with both
independent (diamonds) and joint (stars) schemes.  Observe that independent
predictions scale linearly in both $n$ and $n_p$, but joint predictions are
more costly. Note the change in scale of the $y$-axes from the left to the
right panel. 

\subsection{Competing methodology and software}\label{sec:4compete}

As an alternative to MCMC, others have embraced variational inference (VI) in
which the intractable DGP posterior (\ref{eq:marg}) is approximated with a
simpler family of distributions, often also Gaussian
\citep{damianou2013deep}. Inspired by deep neural networks, \cite{bui2016deep}
proposed an Expectation Propogation (EP) scheme for DGPs, which is closely
related to VI.  \cite{salimbeni2017doubly} broadened previous VI-like
approaches for DGPs by allowing intra-layer dependencies, naming their method
``doubly stochastic variational inference'' (DSVI).  These approaches optimize
rather than integrate, which
requires less work. However, optimization ignores uncertainty;
the fidelity of a VI approximation is linked to the choice of variational
family, rather than directly to computational effort.  More MCMC always
improves posterior resolution; more VI does not.  Hyperparameters
don't neatly fit into variational families otherwise preferred for latent
nodes, so they often get ignored, or their tuning is left to external
validation schemes. By contrast, extra Metropolis easily accommodates a few
more hyperparameters without hassle.

In order to handle training data sizes ($n$) upwards of hundreds of thousands,
VI-based DGPs utilize {\em inducing points}, an umbrella term covering ideas
developed separately as {\em pseudo-inputs} in machine learning
\citep[e.g.,][]{snelson2006} and as {\em predictive processes} in
geostatistics \citep[e.g.,][]{banerjee2008}. Inducing points impose a low-rank
kernel structure by measuring distance-based correlations through a smaller
subset of $m \ll n$ reference locations or ``knots" in $d$-dimensions.
Woodbury matrix identities improve decomposition of the implied full $n \times
n$ structure from $\mathcal{O}(n^3)$ to $\mathcal{O}(nm^2)$.  Although often framed as
an ``approximation'', inducing points can represent a fundamental change to
the underling kernel structure.  Large $n$ and $m$ small enough to
sufficiently speed up calculations can result in low-fidelity or ``blurry'' GP
approximations \citep{wu2022variational}.  Moreover, optimizing inducing point
placement can be fraught with challenges \citep[e.g.,][]{garton2020}. DSVI 
uses $k$-means to place inducing points near clusters of inputs, but 
computer experiments often deploy space-filling designs which would ensure 
there are no clusters.

The Vecchia approximation offers an alternative to inducing points
without introducing auxiliary quantities.  Although it is sometimes cast as a
novel modeling framework rather than an approximation
\citep{datta2016hierarchical}, a key advantage is that it doesn't
fundamentally change the underlying kernel structure -- at least not in the
way inducing points do.  Rather, it more subtly imposes sparsity in its
inverse Cholesky factor. Although Vecchia can be higher on the computational
ladder ($\mathcal{O}(nm^3)$), it is able to provide good approximations with
$m$ much smaller than that required of inducing points without the ``blur'' or
hassle in tuning the locations of $m \times d$ quantities.
\citet{wu2022variational} entertain Vecchia in lieu of inducing points for
ordinary GPs via VI with favorable results. It may only be a matter of time
before Vecchia is deployed with VI for DGPs. We prefer MCMC for its UQ
properties.

Other alternatives to inducing points in a VI context have been suggested,
including random feature expansion \citep[RFE;][]{lazaro2010sparse}.  
Extending RFE from ordinary to DGPs has been the subject of
several recent papers
\citep{cutajar2017random,laradji2019efficient,marmin2022deep}, with some
success.  Others have taken the opposite route, keeping inducing points but
swapping out VI for Hamiltonian Monte Carlo
\citep[HMC;][]{betancourt2017conceptual} for DGPs \citep{havasi2018inference}.
HMC has an advantage over VI in that hyperparameters can easily be subsumed
into the inferential apparatus without external validation.   We show [Section
\ref{sec:5}] that this leads to performance gains on predictive accuracy in
surrogate modeling settings. Nevertheless, DSVI is generally considered the
state-of-the-art for DGPs in machine learning.  We think
this is largely to do with computation and prowess in classification tasks.
DSVI's inducing point approximation enables mini-batching to massively
distribute a stochastic gradient descent.  This seems to work well
in classification settings where resolution drawbacks are less acute, but
our experience [Section \ref{sec:5}] suggests this may not extend
to the low-noise regression settings encountered in surrogate
 modeling of computer simulations.

Ultimately, benchmarking against such alternatives comes down to software, as
even the best methodological ideas are only as good, in practice, as their
implementations.  DSVI is neatly packaged in {\tt gpflux} for {\sf Python}
\citep{dutordoir2021gpflux}, but requires specifying hyperparameters. Default
settings were not ideal for our test problems, and we found manual tuning to
be cumbersome.  The sampling-based HMC implementation is
available on the authors' GitHub page \citep{havasi2018inference}.  This
performs better, we think precisely because of its ability to more
automatically tune hyperparameters in an Empirical Bayes/EM fashion, but
uncertainty in these is not included in the posterior predictions. RFE
software is on the authors' GitHub page \citep{cutajar2017random}, but relies
on the specification of myriad inputs, with few defaults provided. Despite
attempts to port example uses to our surrogate modeling setting, we had
limited success.  We found that results were uniformly inferior to DSVI, and
no predictive uncertainties were provided leading us to ultimately drop RFE
from further consideration.  Finally, EP is available on the authors' GitHub
page \citep{bui2016deep}, but was written in {\sf Python2} and relies on
legacy versions of several dependencies; we were unable to reproduce a suitable
environment to try their code.

\section{Empirical results}\label{sec:5}

Here we entertain MC benchmarking exercises on two simulated examples and one
real-world computer experiment.  Code to re-produce all results (including for
competitors) is available on our public GitHub repository:
\url{https://bitbucket.org/gramacylab/deepgp-ex/}. We include the following
comparators:
\begin{itemize}
	\item DGP VEC: our Vecchia-DGP,
	via {\tt deepgp} using defaults, Mat\`ern $\nu = 5/2$ kernel,
	independent predictions, and ``warped'' conditioning sets.
	See Section \ref{sec:4software}.
	\item DGP FULL: full, un-approximated analog of DGP VEC, with everything
	otherwise identical (also via {\tt deepgp}).  This comparator was not
	feasible for all data sizes.
	\item DGP DSVI: from \cite{salimbeni2017doubly}, implemented in {\tt
	gpflux}, with Mat\`ern $\nu = 5/2$ kernel.  We follow package examples
	in using 100 $k$-means located inducing points.  For numerical stability we
	required {\tt eps = 1e-4}, lower bounding the noise parameter.
	\item DGP HMC: from
	\cite{havasi2018inference}, again using 100 inducing points.  This code only
	supports a squared exponential kernel and estimates (i.e., does not fix) the
	noise parameter ($g$/{\tt eps}).  We found no easy way to adjust
	these specifications, so we let them be.
	\item \added{GP: stationary un-approximated Gaussian process following 
	Eq.~(\ref{eq:gppred}) with Mat\'ern $\nu = 5/2$ kernel and anisotropic lengthscales 
	estimated through MLE.  This comparator was not feasible for all data sizes.}
	\item GP SVEC: scaled Vecchia-GP of \cite{katzfuss2020scaled}, 
	via {\tt GPvecchia} and {\tt GpGp}. This is a fast ``shallow'' GP where
	kernel hyperparameters are estimated (\ref{eq:veclik}) via
	lengthscale-adjusted conditioning sets.  We use {\tt m = 25}, Mat\'ern $\nu
	= 5/2$ kernel, and independent predictions to match DGP VEC.
\end{itemize}
All DGP variations are restricted to two layers. Our metrics include
out-of-sample root mean squared error (RMSE) and continuous rank probability
score \cite[CRPS;][]{gneiting2007strictly}.  \added{Definitions are provided in
Supp.~\ref{app:metrics}.} Lower is better for both. While
RMSE focuses on accuracy of predictive means, CRPS incorporates point-wise
predictive variances, thus providing insight into UQ. Although our DGP VEC is
able to provide full predictive covariance, our competitors DGP DSVI, DGP HMC,
and GP SVEC cannot. \added{Computation times are reported in Supp.~\ref{app:time}.}

\subsection{Simulated examples}\label{sec:5sims}

\paragraph{Schaffer Function.} The two-dimensional ``fourth'' Schaffer function
\added{(Supp.~\ref{app:schaffer})} can be found in the Virtual Library of 
Simulation Experiments
\citep[VLSE;][]{surjanovic2013virtual}.  We follow the second variation
therein using $X\in[-2,2]^2$. The function 
is characterized by steep curved inclines followed by immediate drops. These
quick turns are challenging for stationary GPs, making the Schaffer function
an excellent candidate for DGPs.  We fit models to Latin hypercube samples
\citep[LHS;][]{mckay1979comparison} of training sizes $n \in \{100, 500,
1000\}$ with fixed noise $g = 10^{-8}$.  We use LHS testing sets of size $n_p
= 500$.

\begin{figure}[ht!]
\centering
\includegraphics[width=18cm,trim=10 10 0 10]{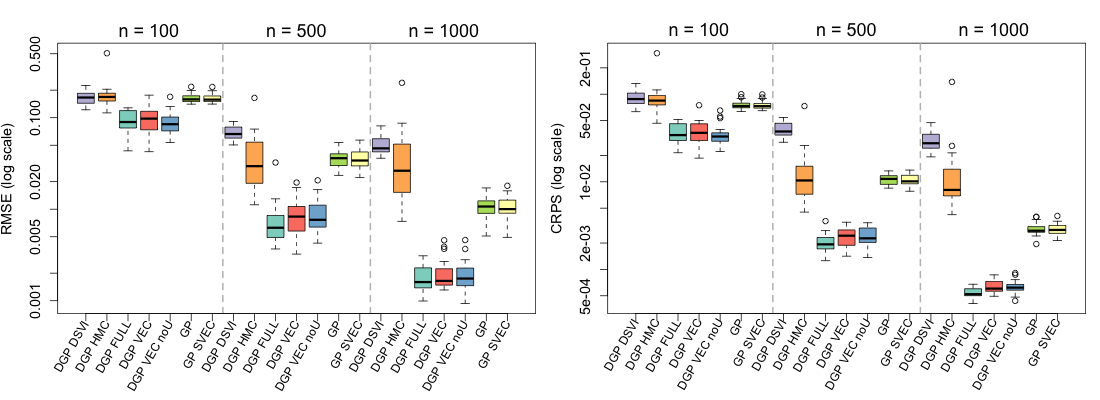}
\caption{RMSE (left) and CRPS (right) on log scales for fits to the 
2d Schaffer function as training size ($n$) increases.  
Boxplots represent the spread of \added{20 MC repetitions.}}
\label{fig:schaffer_results}
\end{figure}

Results for \added{20 MC repetitions} -- all stochastic components
from training/testing re-randomized -- are displayed in Figure
\ref{fig:schaffer_results}.
All variations of our DGP fits outperform competitors by both metrics.
Crucially, our Vecchia-DGP (DGP VEC) matches the performance of the
un-approximated DGP (DGP FULL). For this example we additionally implemented a
``DGP VEC noU'' comparator, identical to DGP VEC but without
ordering/conditioning sets reset after pre-burn-in. Observe that this
additional work is of dubious benefit empirically.  Aside from MC variability,
DGP VEC matches DGP VEC noU. 
Going forward we shall drop DGP VEC noU, focusing on our preferred DGP VEC
setup, without evidence that results are better or worse and despite the
additional (marginal) cost. Additional discussion is deferred to Section
\ref{sec:6}.  Finally, at the outset we expected the stationary GPs 
(\added{GP and} GP SVEC) to
perform poorly given the complexity of the response surface, but were
surprised to see them holding their own against DGP HMC, and eventually
surpassing it with $n=1000$.  We suspect this is a consequence of ``blurry''
inducing point approximations.  \added{The Vecchia approximated GP (GP SVEC) matched 
the performance of the full GP.  Further study of the choice
of conditioning set size $m$ for both GPs and DGPs is provided in 
Supp.~\ref{app:m}.}

\begin{figure}[ht!]
\centering
\includegraphics[width=18cm,trim=10 30 0 10]{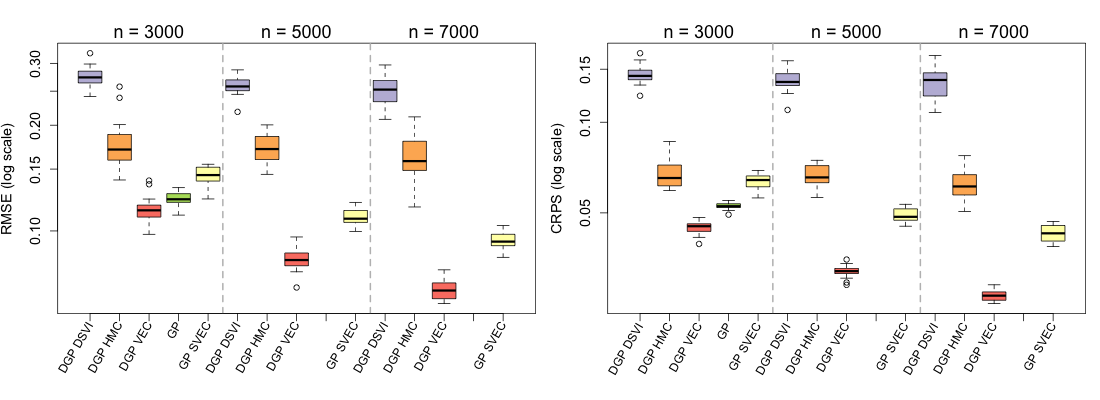}
\caption{RMSE (left) and CRPS (right) on log scales for the 
4d G-function.}
\label{fig:g_results}
\end{figure}

\paragraph{G-function.} The G-function 
\citep[][\added{Supp.~\ref{app:g}}]{marrel2009calculations},
also in the VLSE, is defined in arbitrary dimension.  We worked with
$d = 2$ in Figures \ref{fig:g} and \ref{fig:g_order}.  Here, we expand to
$d = 4$.  Higher dimensionality raises modeling challenges and demands
larger training sets.  We fit models to LHS samples of training sizes $n \in
\{3000, 5000, 7000\}$ with fixed noise $g = 10^{-8}$.  LHS testing sets were
of size $n_p = 5000$. Results for \added{20 MC repetitions} are displayed in Figure
\ref{fig:g_results}.  \added{Variations upon this setup with stochastic 
noise and higher input dimension, yielding similar results, are provided in 
Supp.~\ref{app:gextra}.} 

Again, our DGP VEC outperforms both deep and shallow competitors. DGP HMC,
aided by its ability to estimate hyperparameters, bests DGP DSVI, but neither
benefits from additional training data.  Their predictive capability appears
to saturate; again, we suspect inducing points are to blame.  While it is
possible to increase the number of inducing points and potentially improve
things, this requires adjustments to the source code and, in our experience,
yields only marginal improvements before computation becomes prohibitive.  GP
SVEC is able to adapt to larger training sizes and surpass these deeper
models. Our DGP VEC benefits from estimation of hyperparameters,
additional depth, and learning from additional training data.  \added{While
the full GP was not feasible for larger sample sizes, it significantly
outperformed its Vecchia approximated GP counterpart but was not able to catch
the non-stationary DGP VEC}.

\subsection{Satellite drag simulation}\label{sec:5satellite}

The {\it Test Particle Monte Carlo (TPM)} simulator \citep{mehta2014modeling}
models the bombardment of satellites in low earth orbit by atmospheric
particles, returning coefficients of drag based on particle
composition and input variables specifying satellite configuration.
Researchers at Los Alamos National Lab, who developed TPM,
wished to build a surrogate achieving less than 1\% prediction error (measured
in root mean squared percentage error, RMSPE, \added{Supp.}~\ref{app:metrics}) 
with as few runs of the
simulator as possible. \cite{sun2019emulating} used locally approximated GPs
to reach the 1\% goal with one million training data points.  Later,
\cite{katzfuss2020scaled} achieved lower RMSPE's using scaled
Vecchia-GPs (GP SVEC). Here, we show that our Vecchia-DGP is able to beat
the 1\% RMSPE benchmark with as few as $n = 50,000$ (and can beat it
consistently with $n = 100,000$), and provide better UQ than the stationary GP
SVEC alternative.

We work specifically with the GRACE satellite, specified by a
seven-dimensional input configuration, and a pure Helium atmospheric
composition. For training data, we use random samples from
\cite{sun2019emulating}'s one million runs in sizes of $n \in \{10000, 50000,
100000\}$. We use random out-of-sample testing sets of size $n_p = 50,000$
from the complement, and follow \cite{sun2019emulating} in fixing $g =
10^{-4}$. TPM simulations are technically stochastic, but the noise is very
small.

In light of these large training data sizes and the accompanying computational
burden, we make some strategic choices to initialize our DGP models and set up
the MCMC for faster burn-in.  First, we scale the seven input variables using
estimated vectorized length-scales from  GP SVEC (e.g. $X_i/\sqrt{\theta_i}$).
This ``pre-scaling'' is common in computer surrogate modeling
\citep[e.g.][]{wycoff2021sensitivity}, and mirrors the ``scaled'' component of
the GP SVEC model.  Second, we ``seed'' our MCMC by first running a long,
thoroughly burned-in, set of iterations for one sample of $n = 10{,}000$ and
using the burned-in samples from this fit to initialize the chains for the
larger data sets.  This isn't necessary in practice but helps reduce the
burden of repeated applications in a MC benchmarking context.

\begin{figure}[ht!]
\centering
\includegraphics[width=18cm,trim=10 30 0 20]{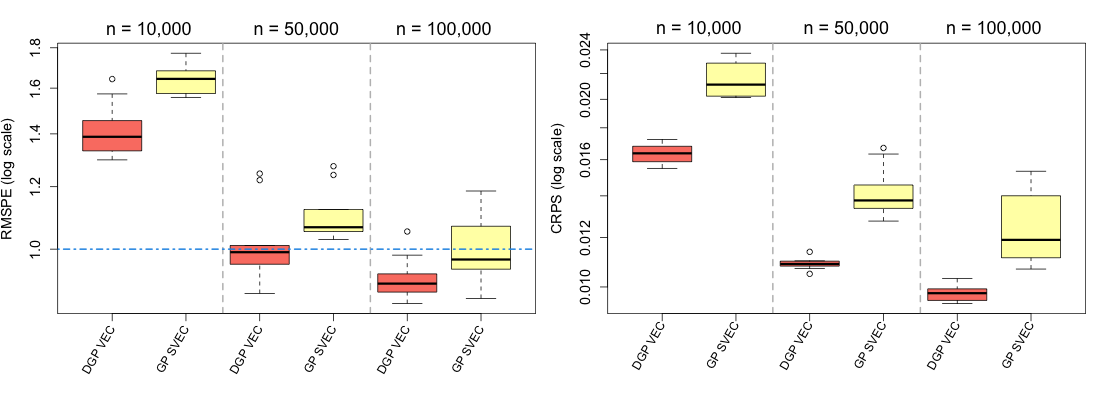}
\caption{RMPSE (left) and CRPS (right) on log scales for fits to the satellite
drag simulation. DGP DSVI and DGP HMC are omitted (with RMSPE's between
30-35\%). Horizontal line marks 1\% RMSPE goal.}
\label{fig:satellite_results}
\end{figure}

Results for 10 MC repetitions are displayed in Figure
\ref{fig:satellite_results}.  Observe that DGP DSVI and DGP HMC results are
omitted from these plots.  Because
they were not competitive (each producing RMSPE's between 30-35\%), their 
inclusion would severely expand the $y$-axis of the plots and render
them hard to read. Our DGP VEC consistently outperforms the
shallow/stationary GP SVEC and is able to achieve the 1\% RMSPE goal with as
few as 50,000 training observations.  Compared to
GP SVEC counterparts (with matched training/testing data), DGP VEC models 
have lower CRPS in all 30 MC repetitions (represented by 3
boxplots of 10) and lower RMSPE in 28 of the 30.

\section{Discussion}\label{sec:6}

In this work we have extended the capabilities of full posterior inference for
deep Gaussian processes (DGPs) through the Vecchia
approximation.  DGPs offer more flexibility than shallow GPs as they
are able to accommodate non-stationarity through the warpings of latent
Gaussian layers.  But inference is slow owing to cubic matrix
decomposition. With a Vecchia approximation at each Gaussian layer, our
fully-Bayesian DGP enjoys computational costs scaling linearly-in-$n$. We
demonstrated the superior UQ and predictive power of Vecchia-DGPs over
both approximate DGP competitors and stationary GPs.

We envision many opportunities for extension.  Most notably, we restricted our
simulation studies to those where the noise parameter ($g$) was fixed to a
small value.  While many computer simulations are deterministic, others are
increasingly stochastic in nature \citep{baker2022analyzing} and prompt
estimation of this hyperparameter, and possibly others.  Our {\tt deepgp}
software is capable of estimating $g$ through additional Metropolis steps. In
the Vecchia context, our implementation incorporates the $g$ parameter
directly into the kernel by adding it to the diagonal of $\Sigma(x_i)$ and
$\Sigma(X_{c(i)})$ in Eq.~(\ref{eq:vecdirect}). Thus $g$ is subsumed into
$B_i(W)$ and $\sigma_i^2(W)$ which in turn populate $U_w$ (\ref{eq:u}).  As a
demonstration of this capability, we provide a noisy simulated example in
Supp.~\ref{app:gextra}.
Results are very similar to those of Figure \ref{fig:g_results}; our Vecchia-DGP 
outperforms across the board.  Some authors separate responses into ``true/latent'' and
``actual/observed'' variables and caution against incorporating the noise
parameter directly into the kernel
\citep[e.g.][]{katzfuss2021general,datta2016hierarchical}, but we have not
seen any drawbacks to our simplified approach in the DGP context. Furthermore,
it may be possible to extend our DGP model to accommodate heteroskedastic
noise through additional latent Gaussian random variables
\citep{binois2018practical}, although we caution that a model that is too
flexible may fall prey to a signal--noise identifiability trap.  One
remedy is to provide for replicates in the design
\citep{binois2019replication}.

We restricted our empirical studies to a conditioning set size of $m = 25$,
after finding limited advantage to larger $m$. \added{[Supp.~\ref{app:m}
explores varying $m$.]}  
This echoes other GP-Vecchia
works which have found success with small conditioning sets
\citep[e.g.][]{datta2016hierarchical,katzfuss2020scaled,wu2022variational,stroud2017bayesian}.
\added{However, the size of $m$ may be more important
in larger dimensional problems.  It may also be advantageous to use larger
$m$ for prediction, as it is otherwise rather cheaper than training}.
Similarly, we entertained only two-layer DGPs.  Our experience with
three-layer DGPs for surrogate modeling is that one of the
latent layers settles into a near identity mapping, resulting in a lot more
computation for little gain.  In some cases, the added flexibility of a
three-layer DGP can lead to overfitting and adversely affect predictions/UQ
\citep{sauer2021active}.  Our test functions and real-data computer
simulations may not be non-stationary enough to warrant two or more levels of
warping.

Perhaps the most intriguing extension lies in the choice of the Vecchia
ordering and conditioning sets.  Common practice, as we embraced, involves
simply fixing an ordering and conditioning, sometimes informed by the data as
when NNs are scaled or warped.  Many works have investigated the effects of
different ordering/conditioning structures
\citep[e.g.][]{katzfuss2021general,guinness2018permutation,stein2004approximating},
yet these analyses have all been {\it post hoc}.  If we view the ordering and
conditioning structure as components of the stochastic (data generating)
process, there may be potential to {\it learn} these, either through
maximization or similar MCMC sampling.  In our simulation studies, updating
the NN conditioning sets based on learned latent warpings did not affect
posterior predictions, perhaps suggesting that inference for these would have
similar null-effects.  Yet we suspect this is because the learned latent
warpings we have encountered tend to rotate and stretch inputs, resulting in
minimal effects on the proximity of observations.  We are continuing to
``hunt'' for examples where more flexible neighborhoods are valuable.

\subsection*{Acknowledgements}

This work was supported by the U.S. Department of Energy, Office of
Science, Office of Advanced Scientific Computing Research and Office of
High Energy Physics, Scientific Discovery through Advanced Computing
(SciDAC) program under Award Number 0000231018.

\bibliographystyle{jasa}
\bibliography{references}

\newpage
\begin{center}
{\large\bf SUPPLEMENTARY MATERIAL}
\end{center}
\appendix

\section{Simulation functions}

\subsection{Schaffer function}\label{app:schaffer}

\added{
The two-dimensional ``fourth'' Schaffer function \citep{surjanovic2013virtual}
is defined as 
\[
f(x_1, x_2) = 0.5 + \frac{\cos^2\left(\sin(|x_1^2 - x_2^2|)\right) - 0.5}
	{\left[1 + 0.001\left(x_1^2 + x_2^2\right)\right]^2}.
\]
We use the restricted domain $X\in[-2, 2]^2$.}

\subsection{G-function}\label{app:g}

\added{
The G-function \citep{marrel2009calculations} is defined in $d$-dimension over
the unit cube $X\in[0, 1]^d$ as
\[
f(\mathbf{x}) = \prod_{i=1}^d \frac{|4x_i - 2| + a_i}{1 + a_i}
\quad\textrm{where}\quad a_i = \frac{i - 2}{2}
\quad\textrm{for all}\quad i = 1, \dots, d.
\]}

\section{Performance metrics}\label{app:metrics}

\subsection{Prediction error}

\added{
Let $\mu^\star$ represent posterior mean predictions at $n_p$ testing
locations.  Let $y^\mathrm{true}$ represent the corresponding observed 
values. Root mean squared error (RMSE) and root mean squared
prediction error (RMSPE) are respectively defined as
\[
\mathrm{RMSE} = \sqrt{\frac{1}{n_p}\sum_{i = 1}^{n_p} 
	\left(\left(\mu^\star_i - y^\mathrm{true}_i\right)^2\right)}
\quad\quad
\mathrm{RMSPE} = \sqrt{\frac{1}{n_p}\sum_{i = 1}^{n_p}
	\left(\left(100*\frac{\mu^\star_i - y^\mathrm{true}_i}{y^\mathrm{true}_i}\right)^2\right)}
\]}

\subsection{Uncertainty quantification}

\added{
Given a Gaussian posterior predictive distribution with predicted 
mean $\mu^\star$ and point-wise standard deviations $\sigma^\star$,
the continuous rank probability score \citep[CRPS;][]{gneiting2007strictly}
is defined as
\[
\mathrm{CRPS}\left(y^\mathrm{true} \mid \mu^\star, \sigma^\star\right) = 
\frac{1}{n_p}\sum_{i=1}^{n_p}\left[
	\sigma_i^\star \left(\frac{1}{\sqrt{\pi}} - 
	2\phi(z_i) - z_i\left(2 * \Phi(z_i) - 1 \right)\right)\right]
\quad\textrm{for}\quad 
z_i = \frac{y_i^\mathrm{true} - \mu_i^\star}{\sigma_i^\star}
\]
where $\phi$ is the standard Gaussian pdf and $\Phi$ is the standard
Gaussian cdf.}

\section{Derivations}

\subsection{Partitioned matrix inverse}\label{app:inv}

The inverse of a partitioned matrix follows \citep{barnett1979matrix}:
\[
\begin{bmatrix} A_{11} & A_{12} \\ A_{21} & A_{22} \end{bmatrix} = 
\begin{bmatrix} B_{11} & B_{12} \\ B_{21} & B_{22} \end{bmatrix}^{-1} 
\quad\textrm{where}\quad
\begin{array}{rl}
A_{11} &= \left(B_{11} - B_{12}B_{22}^{-1}B_{21}\right)^{-1} \\
A_{12} &= -B_{11}^{-1}B_{12}\left(B_{22} - B_{21}B_{11}^{-1}B_{12}\right)^{-1} \\
A_{21} &= -B_{22}^{-1}B_{21}\left(B_{11} - B_{12}B_{22}^{-1}B_{21}\right)^{-1} \\
A_{22} &= \left(B_{22} - B_{21}B_{11}^{-1}B_{12}\right)^{-1} \\
\end{array}
\]

\subsection{Vecchia posterior predictive moments}\label{app:pred}

We aim to predict $\mathcal{Y}$ at locations $\mathcal{W}$ conditioned on 
observed $Y$ and $W$.  We assume a zero-mean Gaussian process prior distribution,
\[
\begin{bmatrix} Y \\ \mathcal{Y} \end{bmatrix}
\sim \mathcal{N}_{n + n_p} \left(0, \;\; \Sigma_{\textrm{stack}}\right)
\quad\textrm{where}\quad 
\Sigma_\textrm{stack} = \Sigma\left(\begin{bmatrix} W \\ \mathcal{W} \end{bmatrix}\right) = 
\begin{bmatrix} \Sigma(W) & \Sigma(W, \mathcal{W}) \\ 
	\Sigma(\mathcal{W}, W) & \Sigma(\mathcal{W}) \end{bmatrix}.
\]
Under the Vecchia-approximation, we decompose the precision matrix using the sparse
upper-lower Cholesky decomposition, with entries populated according to (\ref{eq:u}),
\[
U_{\textrm{stack}} = \begin{bmatrix} U_{w} & U_{w,\mathcal{W}} \\
	0 & U_{\mathcal{W}} \end{bmatrix}
\quad\textrm{such that}\quad
\Sigma_{\textrm{stack}} = \left(U_{\textrm{stack}} U_{\textrm{stack}}^\top\right)^{-1}
= \left(\begin{bmatrix}
U_wU_w^\top + U_{w,\mathcal{W}}U_{w,\mathcal{W}}^\top & U_{w,\mathcal{W}}U_\mathcal{W}^\top \\
U_\mathcal{W}U_{w,\mathcal{W}}^\top & U_\mathcal{W}U_\mathcal{W}^\top \\
\end{bmatrix}\right)^{-1}.
\]
We aim to find a closed-form solution to the posterior predictive moments (\ref{eq:gppred}),
\[
\mathcal{Y} \mid Y, W \sim \mathcal{N}_{n_p}\left(\mu^\star, \Sigma^\star \right) 
\quad\textrm{for}\quad 
\begin{array}{rl}
\mu^\star &= \Sigma(\mathcal{W}, W)\Sigma(W)^{-1} Y \\
\Sigma^\star &= \Sigma(\mathcal{W}) - \Sigma(\mathcal{W},W)\Sigma(W)^{-1}\Sigma(W, \mathcal{W}), \\
\end{array}
\]
which can avoid dense covariance matrices by instead relying on elements of the sparse 
$U_\textrm{stack}$.  The simplification of $\Sigma^\star$ follows directly from the 
partitioned matrix inverse (App.~\ref{app:inv}),
\[
U_\mathcal{W}U_\mathcal{W}^\top = \left(\Sigma(\mathcal{W}) - \Sigma(\mathcal{W},W)
	\Sigma(W)^{-1}\Sigma(W, \mathcal{W})\right)^{-1}
\quad\Longrightarrow\quad \Sigma^\star = \left(U_\mathcal{W}U_\mathcal{W}^\top\right)^{-1}.
\]
The calculation of $\mu^\star$ first involves simplification of $\Sigma(W)$ and 
$\Sigma(\mathcal{W}, W)$, again using partitioned matrix inverses (App.~\ref{app:inv}):
\[
\begin{aligned}
\Sigma(W) &= \left(U_wU_w^\top + U_{w,\mathcal{W}}U_{w,\mathcal{W}}^\top
 - U_{w,\mathcal{W}}U_\mathcal{W}^\top \left(U_\mathcal{W}U_\mathcal{W}^\top\right)^{-1}
 U_\mathcal{W}U_{w,\mathcal{W}}^\top\right)^{-1} \\
 &= \left(U_wU_w^\top\right)^{-1} \\
\Sigma(\mathcal{W}, W) &= -\left(U_\mathcal{W}U_\mathcal{W}^\top\right)^{-1}U_\mathcal{W}
	U_{w,\mathcal{W}}^\top\left(U_wU_w^\top + U_{w,\mathcal{W}}U_{w,\mathcal{W}}^\top - 
	U_{w,\mathcal{W}}U_\mathcal{W}^\top\left(U_\mathcal{W}U_\mathcal{W}^\top\right)^{-1}
	U_\mathcal{W}U_{w,\mathcal{W}}^\top\right)^{-1} \\
&= -\left(U_\mathcal{W}U_\mathcal{W}^\top\right)^{-1}U_\mathcal{W}
	U_{w,\mathcal{W}}^\top\left(U_wU_w^\top\right)^{-1}.
\end{aligned}
\]
Together, these yield
\[
\begin{aligned}
\mu^\star &= \Sigma(\mathcal{W}, W)\Sigma(W)^{-1} Y \\
	&= -\left(U_\mathcal{W}U_\mathcal{W}^\top\right)^{-1}U_\mathcal{W}
	U_{w,\mathcal{W}}^\top\left(U_wU_w^\top\right)^{-1} U_wU_w^\top Y \\
	&= -\left(U_\mathcal{W}^\top\right)^{-1}U_{w,\mathcal{W}}^\top Y.
\end{aligned}
\]

\section{Conditioning set size}\label{app:m}

\begin{figure}[ht!]
\centering
\includegraphics[width=18cm,trim=10 20 0 10]{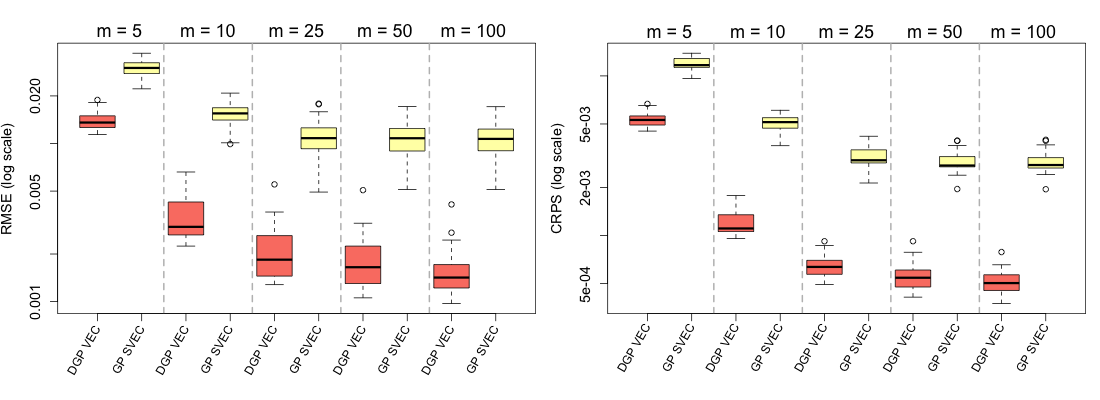}
\caption{\added{RMSE (left) and CRPS (right) for fits to the 2d Schaffer function
as conditioning set size ($m$) increases (same $m$ used for training and prediction).  
Boxplots represent the spread of 20 repetitions.}}
\label{fig:m}
\end{figure}

\added{
Here we evaluate Vecchia-DGP (DGP VEC) and scaled Vecchia-GP (GP SVEC)
models on the two-dimensional Schaffer function while varying the conditioning
set size, $m$.  The set-up is identical to that of Section \ref{sec:5sims}, 
with the exception that we worked with $n=1,000$ and $m\in\{5, 10, 25, 50, 100\}$.  Resulting RMSE and CRPS from 20 MC repetitions are shown in Figure \ref{fig:m}.}

\added{
As expected, increasing the size of the conditioning set improves  
predictive accuracy for both models. However, the 
benefits of conditioning on more points diminishes beyond $m = 25$. 
While present in both models, this ``leveling off'' effect appears more starkly 
in the stationary GP SVEC than the non-stationary DGP VEC.
Table \ref{tab:mtime} reports computation time (in minutes on a
16-core hyperthreaded, Intel i9 CPU at 3.6GHz) for each model/$m$ 
configuration.  The optimization-based inference of the GP SVEC model
is fast enough that larger $m$ do not incur additional computational costs,
practically speaking.
The sampling based nature of DGP VEC reveals the cubic costs
of larger conditioning sets ($\mathcal{O}(nm^3)$).  These cubic
costs, and the minimal predictive improvements realized with larger $m$, 
motivate our choice of $m = 25$ for all other exercises and as the default
setting in our software.}
\begin{center}
\begin{table}[h]
\caption{\added{Computation times in minutes for 2d Schaffer
function exercise of Figure \ref{fig:m}.}}
\begin{tabular}{l | c | c | c | c | c}
{\bf Model} & $m = 5$ & $m = 10$ & $m = 25$ & $m = 50$ & $m = 100$ \\
\hline
DGP VEC & 3.16 & 4.19 & 8.91 & 25.80 & 92.09 \\
\hline
GP SVEC & 0.02 & 0.01 & 0.01 & 0.02 & 0.07 \\
\end{tabular}
\label{tab:mtime}
\end{table}
\end{center}

\section{Additional simulations}\label{app:gextra}

\subsection{Simulation with noise}\label{app:g4noise}
As an example of a noisy simulation, we re-create the Monte Carlo exercise for
the G-function \citep{marrel2009calculations} with the addition of additive
Gaussian noise $\epsilon_i \sim \mathcal{N}(0, 0.01^2)$. The set-up is
equivalent to that of Section \ref{sec:5sims}, except that each model is now
tasked with estimating a noise parameter (i.e. {\tt true\_g = NULL} in {\tt
deepgp}).  The DGP HMC and GP SVEC models have built-in capability to estimate
noise.  The DGP DSVI model does not, so we simply fix the noise parameter to
the true variance ($g = 0.01$).  To account for the extra challenge of
distinguishing signal from noise, we double the data sizes to  $n \in \{6000,
10000, 14000\}$ and $n_p = 10000$.  Code to reproduce these results is
available in our github repository.
\begin{figure}[ht!]
\centering
\includegraphics[width=18cm,trim=10 20 0 10]{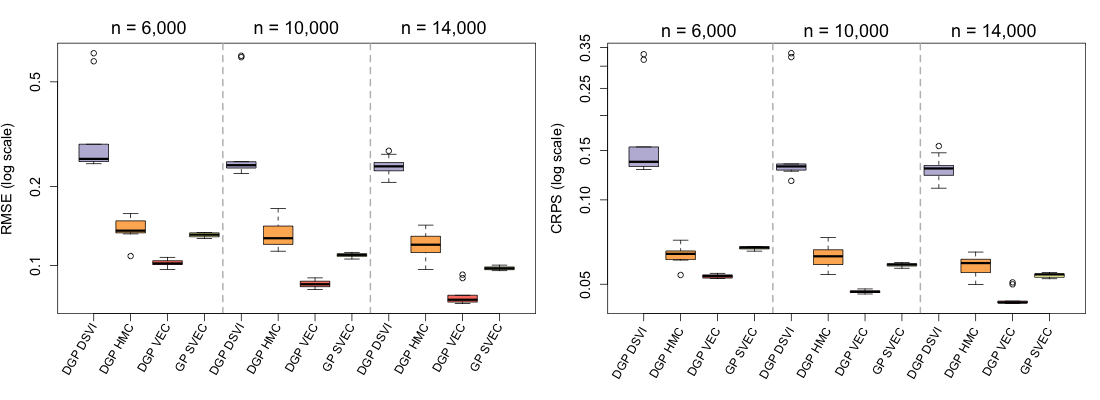}
\caption{RMSE (left) and CRPS (right) on log scales for the 
4d G-function observed with Gaussian white noise.
Boxplots represent the spread of 10 MC repetitions.}
\label{fig:g4_noise}
\end{figure}

The results resemble Figure \ref{fig:g_results}, suggesting the addition of noise
does not affect the comparative efficacy of the models.  DGP HMC and GP SVEC perform 
similarly, the former benefiting from the flexibility of DGP layers and the latter 
benefiting from the Vecchia approximation (as compared to inducing points).  
The DGP VEC model outperforms across the board.  When 
matched by training/testing data, DGP VEC had lower RMSE and CRPS than 
each of these comparators in 30/30 trials. 

\subsection{Higher dimensional simulation}\label{app:g6}
\added{To display functionality in higher dimension, we again re-create the Monte
Carlo exercise for the G-function, this time expanding to $d=6$.  The set-up
is equivalent to that of Section \ref{sec:5sims}.  Higher dimension
demands larger data sizes; we entertain random LHS training designs of size 
$n \in \{10000, 20000, 30000\}$ and LHS testing designs of size $n_p = 20000$.}
\begin{figure}[ht!]
\centering
\includegraphics[width=18cm,trim=10 20 0 10]{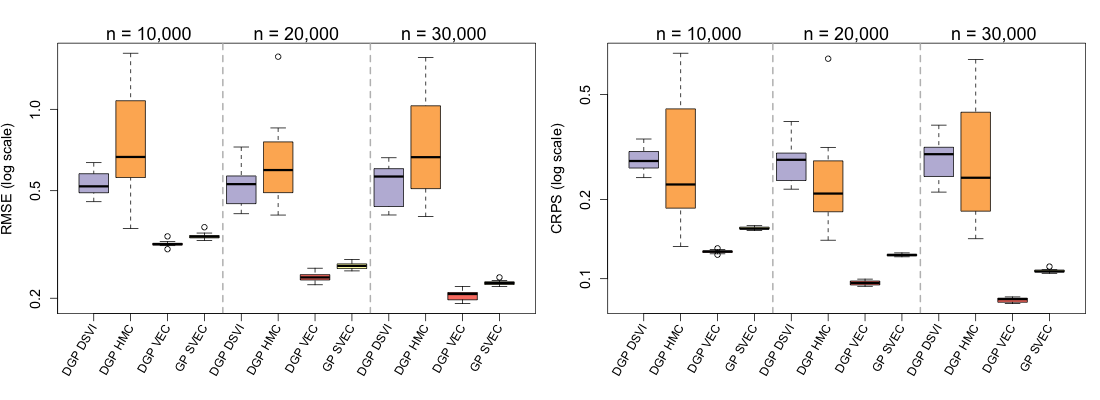}
\caption{\added{RMSE (left) and CRPS (right) on log scales for the 
6d G-function.  Boxplots represent the spread of 10 MC repetitions.}}
\label{fig:g6}
\end{figure}

\added{As in the lower dimensional exercises, DGP VEC outperforms on both prediction
error and uncertainty quantification.  Both DGP DSVI and DGP HMC are limited
by inducing point approximations which are especially blurry in high dimensions.
We note that the DGP HMC model is at a slight disadvantage since it estimates
a noise parameter -- the poor fits from this model may be 
over-estimating the noise.}

\newpage 
\section{Computation times}\label{app:time}

\added{
The following tables report computation times (in minutes) for the
simulation exercises of Section 5 and Supp.~\ref{app:gextra}.  We 
report the computation
time for a single Monte Carlo exercise; times for each
randomized trial are similar.  All times were recorded on
a 16-core hyperthreaded, Intel i9 CPU at 3.6GHz.  The compute times
of MCMC methods are directly tied to the number of MCMC samples collected.  
DGP VEC models sampled 3000 iterations for the Schaffer function, 3000 for 
the G-function, and 2000 for the satellite simulation.  DGP HMC models
utilized the default of 10,000 iterations.
Un-approximated models (DGP FULL and GP) were not run for larger
data sizes due to the cubic computational costs.}

\begin{center}
\begin{table}[ht!]
\caption{\added{Computation times in minutes for the Schaffer function 
exercises of Section \ref{sec:5sims} (Figure \ref{fig:schaffer_results}).}}
\begin{tabular}{l | c | c | c}
{\bf Model} & {\bf N = 100} & {\bf N = 500} & {\bf N = 1,000} \\
\hline
DGP DSVI & 0.17 & 0.50 & 0.97 \\
\hline
DGP HMC & 0.58 & 1.45 & 2.43 \\
\hline
DGP FULL & 1.71 & 194.61 & 1605.94 \\
\hline
DGP VEC & 2.08 & 5.12 & 8.16 \\
\hline
DGP VEC noU. & 1.95 & 5.04 & 8.27 \\
\hline
GP & 0.03 & 0.55 & 3.48 \\
\hline
GP SVEC & 0.04 & 0.08 & 0.07\\
\end{tabular}
\end{table}
\end{center}

\begin{center}
\begin{table}[ht!]
\caption{\added{Computation times in minutes for the G-function 
exercises of Section \ref{sec:5sims} (Figure \ref{fig:g_results}).}}
\begin{tabular}{l | c | c | c}
{\bf Model} & {\bf N = 3,000} & {\bf N = 5,000} & {\bf N = 7,000} \\
\hline
DGP DSVI & 3.10 & 5.11 & 7.14 \\
\hline
DGP HMC & 7.59 & 12.21 & 17.35 \\
\hline
DGP VEC & 39.70 & 64.38 & 89.06 \\
\hline
GP &  807.33 & N/A & N/A \\
\hline
GP SVEC & 0.05 & 0.09 & 0.09 \\
\end{tabular}
\end{table}
\end{center}

\begin{center}
\begin{table}[ht!]
\caption{\added{Computation times in minutes for the satellite simulation
exercises of Section \ref{sec:5satellite} (Figure \ref{fig:satellite_results}).}}
\begin{tabular}{l | c | c | c}
{\bf Model} & {\bf N = 10,000} & {\bf N = 50,000} & {\bf N = 100,000} \\
\hline
DGP DSVI & 12.00 & 59.13 & 109.20 \\
\hline
DGP HMC & 26.72 & 26.29 & 26.47 \\
\hline
DGP VEC & 190.57 & 809.75 & 1606.15 \\
\hline
GP SVEC & 0.17 & 0.18 & 0.18 \\
\end{tabular}
\end{table}
\end{center}

\begin{center}
\begin{table}[ht!]
\caption{\added{Computation times in minutes for the noisy G-function
exercise of Supp.~\ref{app:g4noise} (Figure \ref{fig:g4_noise}).}}
\begin{tabular}{l | c | c | c}
{\bf Model} & {\bf N = 6,000} & {\bf N = 10,000} & {\bf N = 14,000} \\
\hline
DGP DSVI & 6.15 & 10.17 & 14.22 \\
\hline
DGP HMC & 14.69 & 24.94 & 25.19 \\
\hline
DGP VEC & 84.85 & 132.41 & 179.44 \\
\hline
GP SVEC & 0.09 & 0.09 & 0.09 \\
\end{tabular}
\end{table}
\end{center}

\begin{center}
\begin{table}[ht!]
\caption{\added{Computation times in minutes for the 6d G-function
exercise of Supp.~\ref{app:g6} (Figure \ref{fig:g6}).}}
\begin{tabular}{l | c | c | c}
{\bf Model} & {\bf N = 10,000} & {\bf N = 20,000} & {\bf N = 30,000} \\
\hline
DGP DSVI & 11.73 & 21.82 & 35.04 \\
\hline
DGP HMC & 25.91 & 26.05 & 26.14 \\
\hline
DGP VEC & 235.49 & 425.43 & 623.89 \\
\hline
GP SVEC & 0.43 & 0.34 & 0.34 \\
\end{tabular}
\end{table}
\end{center}

\end{document}